\documentclass[sigconf, nonacm]{acmart}
\usepackage{graphicx}
\graphicspath{{images/}}
\usepackage{blindtext}
\usepackage{subfiles} % Best loaded last in the preamble

\AtBeginDocument{%
  \providecommand\BibTeX{{%
    \normalfont B\kern-0.5em{\scshape i\kern-0.25em b}\kern-0.8em\TeX}}}

\usepackage{todonotes}
\usepackage{listings}

\begin{document}

\title{It's About Time: Incorporating Temporality in Retrieval Augmented Language Models}

\author{Anoushka Gade}
\email{anoushka.gade@sjsu.edu}
%\author{}
\affiliation{%
  \institution{San Jose State University}
  \streetaddress{}
  \city{San Jose}
  \state{CA}
  \country{USA}
  \postcode{}
}

\author{Jorjeta Jetcheva}
\email{jorjeta.jetcheva@sjsu.edu}
\orcid{}
\affiliation{%
  \institution{San Jose State University}
  \streetaddress{}
  \city{San Jose}
  \state{CA}
  \country{USA}
  \postcode{}
}
\renewcommand{\shortauthors}{Gade and Jetcheva, et al.}

\begin{abstract}

The web serves as a global repository of knowledge, used by billions of people to search for information. 
Ensuring that users receive the most relevant and up-to-date information, especially in the presence of multiple versions of web content from different time points remains a critical challenge for information retrieval. This challenge has recently been compounded by the increased use of question answering tools trained on Wikipedia or web content and powered by large language models (LLMs) \citep{chatgpt} which have been found to make up information (or hallucinate), and in addition have been shown to struggle with the temporal dimensions of information.  Even Retriever Augmented Language Models (RALMs) which incorporate a document database to reduce LLM hallucination are unable to handle temporal queries correctly. This leads to instances where RALMs respond to queries such as "Who won the Wimbledon Championship?", by retrieving document passages related to Wimbledon but without the ability to differentiate between them based on how recent they are. 

In this paper, we propose and evaluate, TempRALM, a temporally-aware Retriever Augmented Language Model (RALM) with few-shot learning extensions, which takes into account both semantically and temporally relevant documents relative to a given query, rather than relying on semantic similarity alone.  We show that our approach results in up to 74\% improvement in performance over the baseline RALM model, without requiring model pre-training, recalculating or replacing the RALM document index, or adding other computationally intensive elements.

\end{abstract}


% <concept_id>10002951.10003317.10003338.10003341</concept_id>
% <concept_desc>Information systems~Language models</concept_desc>

% \ccsdesc[500]{Information systems~Language models}



\maketitle

\section{Introduction}

The web serves as an ever-expanding reservoir of real-world knowledge, with textual documents constituting a significant fraction of its content.  Moreover, information changes over time, leading to  updates to existing documents, or the addition of new documents. This leads to multiple versions of information from various time frames to co-exist and grow over time. 
A major challenge in information retrieval is ensuring that users get access to the most relevant and up-to-date knowledge at any time.
This challenge has recently been compounded by the increased use of question answering tools powered by large language models (LLMs), which have gained popularity as a result of the release of chatGPT \citep{chatgpt}.  
LLMs have been shown to absorb and serve immense quantities of information from textual data \citep{Petroni}. 
This information is typically derived from a static snapshot of a large number of documents scraped from the web at a specific point in time. However, real-world information changes continuously, frequently on a daily, hourly or even real-time basis. 
To address the challenge of frequently changing data, language models must expand or update their memory on a regular basis, which involves significant computational, financial and environmental costs.

\begin{figure*}[ht]
\centering
\includegraphics[width=\textwidth]{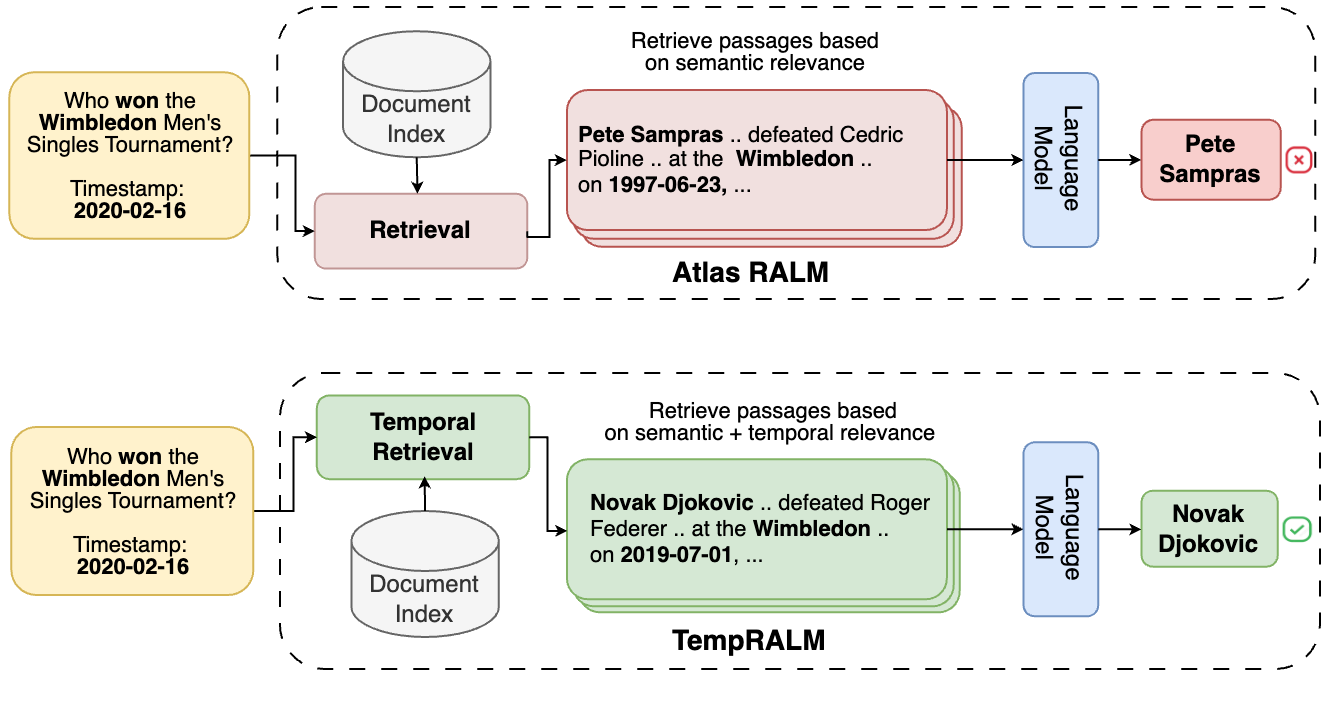}
\caption{Overview of TempRALM: In this figure, we show the difference between Atlas and TempRALM. In TempRALM, the retriever fetches documents on semantic relevance with respect to the query as well as temporal relevance relative to the query-time}
\end{figure*}

In recent years, there has been an increasing interest in using Retrieval Augmented Language Models (RALMs) as a way of addressing the problem of frequently evolving information, and also reducing the tendency of LLMs to make up information known as hallucinations. 
RALMs use an external document corpus as a knowledge source rather than relying solely on parametric memory like standard LLMs.  
This corpus, stored in the form of a document index (a format which makes document retrieval efficient), can be augmented and updated to reflect current versions of documents stored in it, for e.g., Wikipedia documents, and web pages.
%This external corpus can take the form of text documents containing domain-specific information such as Wikipedia, and can be updated as necessary.  
% The RALM architecture is divided into three components: a document index, a retriever that obtains relevant documents from the index, and a language model that generates output. 
For a given text-based query, RALM uses its retriever component to obtain a ranked list of the most relevant documents from the document index and gives them to its language model component (also known as a reader), which uses the ranked documents as context to generates a response to the query. 
% something about the fact that docs may be split into smaller units?
RALMS have been shown to generate more specific and factual responses to queries in knowledge intensive tasks such as question answering and fact checking than stand-alone LLMs \citep{rag_knowledg}. These models have also shown to perform well in few-shot training scenarios \citep{izacard2022atlas} using as few as 64 examples to achieve a good performance. 

Although RALMs perform well in factual question answering, they typically use a document index, which contains a single version of each document. 
However, in many real-world applications, new information continues to be generated, without invalidating or replacing existing information, which results in multiple versions of documents to be present in the document index. 
For example, scientific and medical journals continuously publish new academic papers which are additions and enhancements to the body of knowledge that has been published previously, for e.g., new Natural Language Processing (NLP) research papers are posted to the arxiv platform \cite{arxiv:} almost daily.  
%In order for such scenarios to work in a RALM environment, we would want to preserve both old and new data in the document index. 
This challenge is further compounded when trying to build a question answering system that supports time-sensitive queries such as "Which language model is the state-of-the-art on the latest summarization benchmark?", because standard RALM models would retrieve numerous papers, each stating it has established a new state of the art in document summarization.  This is a direct result of the fact that standard LLMs and the LLM components of RALMs do not take into account temporal metadata \citep{templama}.

In this paper, we show that RALMs face challenges in handling temporality in even simpler and more structured settings. 
% Temporality is an important concept. When we ask a language model to respond to a query at a given point in time, it should provide the most relevant answer at that point in time. 
For example, in the context of frequently changing information such as the identities of the latest winners of the Wimbledon tennis championship, we show that Atlas \citep{izacard2022atlas}, a representative state-of-the-art RALM, is generally unable to provide an answer that is meaningful relative to the time at which the question is asked. For example, if the question "Who won the Wimbledon Men's championship?" is asked on December 31, 2019, current state-of-the-art RALM architectures would retrieve relevant documents by matching the year 2019 in the query timestamp with the Wimbledon match date in the passages. However, if the same question is asked a day later, i.e. on January 1, 2020, the current RALM architecture can no longer retrieve the relevant documents as the year no longer matches, and since they do not have the ability to understand temporal relationships.

% the below doesn't help with our story so have commented it out - we generally don't want to replace information but add anyway because people may ask questions about the data at previous time points, not just what is currently the champions, etc. 
%In a RALM setting, it is not easily possible to keep the document index consistent at all times. Ensuring that the index reflects the most recent information is impossible with datasets such as arxiv we outlined above, because there is no idea of "old" or "new" in such data. 
%Aside from the issue of keeping a consistent index, the cost operations for replacing an index entirely is much higher than adding to it.  

Prior attempts to address this problem have focused on pre-training the language model by inducing a temporal context as plain text \citep{templama}, where the model is trained with a snapshot of timestamp-prefixed data to enable the model to learn to handle temporal data. However, the snapshot it is trained on is static, and pre-training of the models is an expensive operation, which cannot be performed as frequently as real-world data tends to be updated. 
%{\bf *** need more detail here to explain what data is used for training, etc.} 

% \usepackage{float}

In this paper, we introduce a novel, simple, yet effective method to retrieve temporally correct documents with respect to a given query and use it to augment Atlas \citep{izacard2022atlas}, a standard RALM model with few-shot learning extensions. 
%First, we encode a timestamp in each document in our document index, as well as the query, and use few-shot learning to train the model. 
Our model, TempRALM, includes an extension of the RALM retriever's document retrieval and ranking algorithm, to take into account both semantically and temporally relevant documents relative to each query, rather than relying on semantic similarity alone. We show that our approach results in up to 74\% improvement in performance over the baseline Atlas model, while being computationally efficient.  TempRALM does not require pre-training, recalculating or replacing the document index, or the addition of other computationally intensive elements. 

The rest of this paper is organized as follows.  Section~\ref{sec:related-work} overviews related work.  We introduce our methodology and experimental setup in Sections \ref{sec:methodology} and~\ref{sec:experimental-setup}. 
Section~\ref{sec:results} discusses our experimental results.  We conclude the paper with a summary and next steps in Section~\ref{sec:conclusions}.

\begin{figure}[htb]
\centering
\includegraphics[
% width=0.5\textwidth, 
scale=0.7]{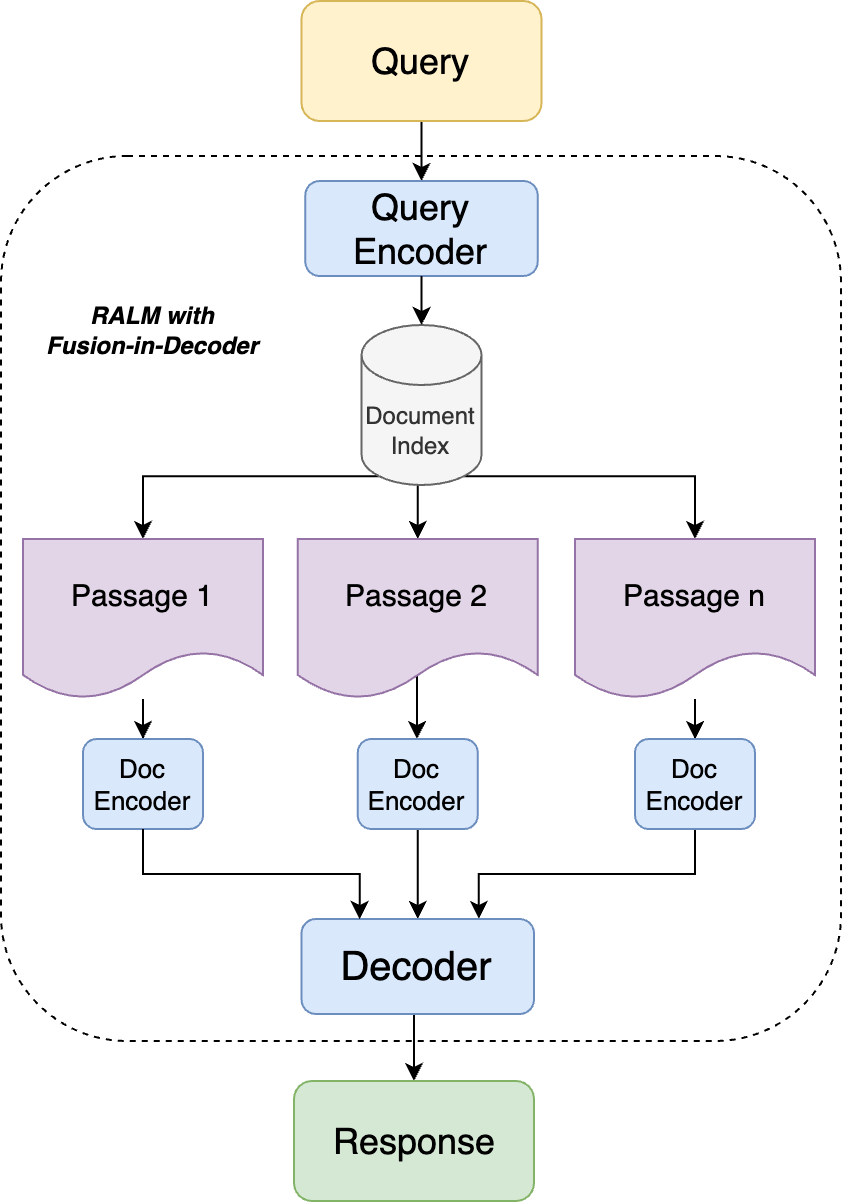}
\caption{Retrieval Augmented Language Model with Fusion-in-Decoder Architecture}
\label{Fig:RAG}
\end{figure}

\section{Related Work}
\label{sec:related-work}

%In Section~\ref{related-work} we introdu....

% \subfile{sections/related_works}
% Writing Related Works here to confirm citations work properly

Large Language Models (LLMs) \citep{brown2020language}, \citep{palm2} and conversational interfaces to these models like ChatGPT \citep{chatgpt}, have the potential to revolutionize our daily lives by helping people discover and understand information in real-time.  However, these systems are prone to hallucinations and highly confident sounding answers that are in fact wrong, limiting their use for knowledge-intensive tasks such as Question-Answering, Fact Checking, etc. Retrieval Augmented Language Models (RALMs)\citep{rag_knowledg} offer a promising approach to mitigating these problems, by adding a document database (also known as document index) as a knowledge source which is used to prompt or contextualize LLM-generated answers. Typically, a Wikipedia snapshot is used to construct the document index of general-purpose RALMs. When a query is made, a set of documents are retrieved (by a retriever model within the RALM) and a response is generated using the query and the retrieved documents as input to an LLM. This approach has been shown to significantly improve the performance of LLMs on knowledge intensive tasks shown by\cite{rag_knowledg}.
Moreover, training the retriever does not require a labeled mapping between queries and documents, and thus does not introduce data labeling complexity to the training process.

To alleviate the need of large datasets for training or fine-tuning RALMs, \citep{izacard2022atlas} propose Atlas, a pre-trained model with few-shot learning capabilities. 
It consists of a pre-trained retriever, \cite{contriever}, which is a dense retriever based on the BERT \cite{BERT} architecture, trained using contrastive loss, as well as a language model, T5, \cite{T5}, trained to perform a range of sequence-to-sequence tasks, e.g., summarization, translation, and question answering. Additionally both models are fine-tuned jointly as a Fusion-in-Decoder \cite{fusion_in_decoder} on common-crawl data using Wikipedia as a document index.  The authors showed that Atlas performs well out-of-box and can be adapted to different tasks such as question-answering, fact-checking, etc with as few as 64 few-shot examples.

An aspect of question answering that has emerged as a challenge for both RALMs and LLMs are questions whose answers evolve over over time as seen in \citep{zhao2022dense}. For example, the answer to "Who is the President of the United States?" will have a different answer every 4 years or so. 

In \citep{templama}, the authors  explore the challenges such questions pose to LLMs, and show how pre-trained models on a single snapshot of the Internet at a given point in time are inherently incapable of answering them correctly. In their work in~\cite{izacard2022atlas}, the authors investigate this problem in the context of RALMs and show that RALMs can handle these questions better than pure LLMs, by replacing the document index with an up-to-date document index. However, this modifications is not sufficient to address question answering in domains such as scientific, medical and legal publications, where it is not easy to replace one fact with another.  In addition, updates to knowledge in these domains result in adding new documents, rather than replacing existing documents with updated versions.

In \citep{situatedqa}, the authors also explore time-sensitive open domain QA, where they provide a Dense Passage Retriever \citep{dpr} with a query modified with additional lingustic context, but see no improvement for temporally dependent questions. They also found that retrieval based models were able to update some world knowledge after swapping the retrieval corpus and fine-tuning with newer data. However, this approach is not efficient to keep models' responses up-to-date. This approach is similar to the approach used in Atlas, where the authors use a static snapshot of Wikipedia, but swap the corpus depending on the temporal context, which is not feasible in scenarios where there are frequent updates in the data.

%This which is inconsistent with the idea of temporally evolving queries, as a static index wont be able to address dynamically evolving questions. 
In \citep{realtimeqa}, the authors proposed to collect temporally evolving questions by sourcing from news websites.
% which as in \citep{templama} they use as a static index which they swap out with updated versions of the documents once they change on their source website. 
They propose a simple solution of a RAG model with a wiki snapshot to address the questions. One limiting factor in this approach is that they use a static index to address dynamically evolving questions.
In our approach, we handle multiple versions of each document, in order to accommodate cases where information is added, without needing to swap out or invalidate prior information.

We build on the the work by \citep{izacard2022atlas} by proposing and evaluating, TempRALM, a novel temporally augmented RALM with few-shot learning capabilities, which can accommodate a document index which contains multiple versions of documents, and does not require model re-training.
%, in order to better address the problem of handling temporally evolving factual questions with a non-consistent index, to show that the retriever can fetch the correct information, even when multiple versions of documents across different points of time co-exist in the document index.

%\section{Problem Formulation}

%\subfile{sections/problem_formulation}

\section{Methodology}

\begin{figure*}[t]
\centering
\includegraphics[width=\textwidth,]{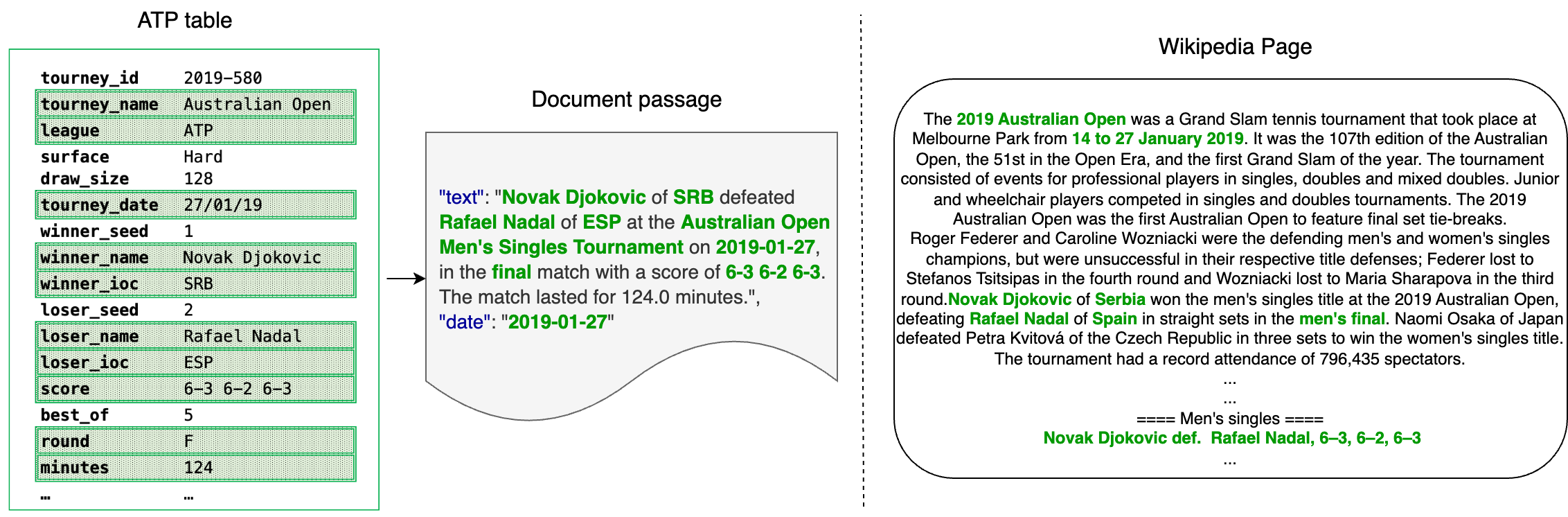}
\caption{Example of tabular data to passage conversion. We convert relevant features of the tabular dataset to textual passages.}
\label{Fig:CSV_to_passage_generation}
\end{figure*}
\label{sec:methodology}
% Problem definition
% Temporal Augmentation
%     1. RALM architecture
%     2. Semantic Score
%     3. TEMPRALM

In our study, we build on the Retrieval Augmented Language Model (RALM) architecture, enhanced with few-shot learning, which was first introduced by the Atlas model~\citep{izacard2022atlas}. Notably, existing RALM approaches, including Atlas, are not able
to take into account the temporal aspects of a query.

To that end, we augment Atlas with a novel temporal retrieval method and evaluate the performance of the resulting model. We overview Atlas and the details of our proposed extensions in the remainder of this section.

\subsection{RALM Overview}
The RALM architecture has three core elements: a document index which contains documents, a retriever to identify the relevant documents (from the document index) for a given query/question, and a language model (also called as reader) that generates the response to the question. 

%For knowledge intensive tasks such as question answering or fact checking, t
The retriever identifies the $top k$ most relevant documents (or passages) for a given query from the (typically very large corpus of) text documents stored in the document index. The language model takes the retrieved documents and the query as input, and generates an answer to the query.

For our document index, we use a time-sensitive dataset that contains text passages that refer to tennis tournament details at different time points, described in Section \ref{sec:experimental-setup}.
We use the pre-trained Atlas generator and retriever components. The retriever is based on the Contriever \citep{contriever}, an information retrieval approach based on continuous dense embeddings. 
The generator is based on a sequence-to-sequence T5 \citep{T5} architecture. It is combined with the retriever with a Fusion-in-Decoder \citep{fusion_in_decoder} modification and language modelling adaptation. 

While we validate our hypothesis using the RALM architecture under few-shot conditions, it is important to note that our approach could be used to extend any retriever that ranks documents using a numerical similarity score between a query and a document corpus. We show the RALM with fusion-in-decoder architecture in Figure \ref{Fig:RAG}

\subsection{Semantic Retrieval Score}
The core computation involved in deciding which documents are relevant to a query is based on computing a semantic similarity-based relevance score between the query and each document in the index and using the dot product of their encoder representations. In particular, for a given query $q$ and document $d$, their encoder representations are obtained independently by encoding them using an encoder $f_\theta$, parameterized by $\theta$, the set of trainable parameters of the encoder. 

% {\bf this model f comes up out of the blue - need to relate to the BERT discussion - isn't there a different encoder for the query vs doc (implied by the text about BERT}

Then the semantic score, $s(q, d)$ between the query $q$ and the document $d$ is computed as the dot product of their resulting encoder representations:

\begin{equation}
    s(q,d) = \langle f_\theta (q), f_\theta (d) \rangle
    \label{eq:semantic}
\end{equation}

% {\bf are we going to call this  "relevance score" or "sematic score" or "sematic similarity score"? need to have consistent naming }

\subsection{Temporal Score} 

% - we need to see if we can have a name for the technique that is more specific and use to reference the proposed approach by name

%For testing our hypothesis, we introduce a new retrieval methodology, that incorporates temporality in its core functionality. 

% {\bf jj: removed subscript in qt and dt because we cannot use the same timestamp t as a subscript for both of them }
We introduce a new method called TempRALM, which augments the retriever with a temporal score $\tau(qt, dt)$, where $qt$ is the timestamp of the query $q$, and $dt$ is the timestamp of document $d$. To calculate $\tau(qt, dt)$,  we take the reciprocal of the time difference between $qt$ and $dt$ to ensure that the temporal score is inversely proportional to the time difference (and thus smaller differences in time result in a larger score), and use a scaling factor $\alpha$ to tune the temporal score's importance relative to the importance of the semantic score.  The temporal score formulation is as follows:

\begin{equation}
    \tau(qt, dt) = \frac{\alpha_{scale}}{qt - dt}\ 
    \label{eq:temporal}
\end{equation} 

To align the temporal score with the numerical range of the semantic score $s(q,d)$ from equation \ref{eq:semantic}, we employ a normalization process. In particular, we subtract from each temporal score, the mean over the temporal scores of all $(q,d)$ pairs, $\mu_\tau$, and divide by the standard deviation over the temporal scores of all $(q,d)$ pairs, $\sigma_\tau$. 
The normalized temporal score is then scaled to the range of the relevance score by multiplying with standard deviation of semantic scores $\sigma_s$ and adding mean of semantic scores $\mu_s$ (both computed over the scores of all $(q,d)$ pairs):

\begin{equation}
    \tau(qt, dt) = \frac{\tau(qt, dt) - \mu_\tau} {\sigma_\tau}\  \times \sigma_s  + \mu_s
    \label{eq:temporal_scaled}
\end{equation} 

\begin{figure*}[t]
\centering
\includegraphics[width=\textwidth,]{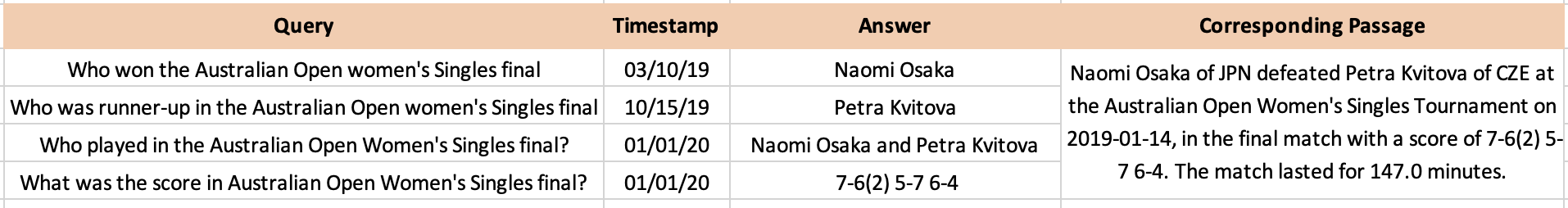}
\caption{Questions asking about the winner, runner-up, finalists and score asked at different timestamps, where the right answer corresponds to the most recent tennis match that occurred before the query timestamp}
\label{Fig:Query_examples}
\end{figure*}

\subsection{TempRALM: A Temporally-Augmented Retriever} 

To incorporate the temporal score shown in equation (\ref{eq:temporal_scaled}) in the retriever, we define an auxiliary retrieval scoring function $TempRet_t$, which is the sum of the semantic score $s(q,d)$ and the temporal score $\tau(qt, dt)$: 

\begin{equation}
    TempRet_t (q, d, qt, dt) = s(q,d) + \tau(qt, dt)\
    \label{eq:retriever_score}
\end{equation} 

% {\bf jj: renamed Ret to TempRet to indicate that this score has a temporal component}

In addition, we extend the retriever by removing passages related to future information with respect to the time of the query.
%\subsection{Masking Future Passages}
In particular, we mask the retrieval score for retrieved passages that have time stamps that are larger than the query timestamp (denoting passages that refer to future information relative to the time of the event specified in the query), to completely eliminate these passages from being considered while ranking the $topk$. We re-write the retrieval computation as follows: 

\begin{equation}
    TempRet_t (q, d, qt, dt)= 
\begin{cases}
    s(q,d) + \tau(qt, dt)\, & \text{if } qt\geq dt\\
    -\infty,              & \text{otherwise}
\end{cases}
\label{eq:final_retriever_score}
\end{equation} 

%\subsubsection{Over Retrieval for more coverage}
Finally, to ensure a comprehensive coverage of relevant passages, we over-retrieve of set of documents from the retriever, and from this over-retrieved set, the $topk$ documents with highest  $TempRet_t$ scores are passed as input to the language model. We conduct an extensive set of experiments to find the optimal number of documents to over-retrieve. 
It's important to note that our extensions do not require re-training any part of the RALM.

% {\bf jj: we need to define what over-retrieving means} 

% By introducing a change in the retriever mechanism, we incorporate temporality through our algorithm, which easily works in conjunction with other retrieval methods, without having to re-train from scratch.

\subsection{Evaluation Metrics}
% % {\bf jj: this is the former problem formulation section, which 
% % fits better here - most of it is actually about the metrics so I renamed it; }

We evaluate the effectiveness of our system on time-specific question answering test sets. % of size 32, 64, and 128. 
We assume a few-shot training set of (query, timestamp, answer) triplets, 

\noindent $Q_{train} = \{(q_1, qt_1, a_1), (q_2, qt_2, a_2), \dots (q_N, qt_N, a_N \}$ and evaluate on a test set of held out queries, timestamps, and answers. 
Our evaluation metric is exact-match between 
%We evaluate the effectiveness of the system on the basis of exact match between 
the predicted and ground-truth answer for each data point in the test set.

\section{Experimental Setup}

\label{sec:experimental-setup}

% \begin{figure*}[t]
% \centering
% \includegraphics[width=\textwidth,]{images/CSV_PAssage.png}
% \caption{Example of tabular data to passage conversion. We convert relevant features of the tabular dataset to textual passages.}
% \label{Fig:CSV_to_passage_generation}
% \end{figure*}

To explore the effectiveness of our temporally augmented RALM model, we formulated a retrieval task based on a use case where information evolves over time. In particular, we curated a dataset based on tennis grand slam data, where 4 major tournaments take place every year and answers to the queries depend on \underline{when} the query is asked, with respect to the timing of the tournament. We describe the details of the model configuration, our dataset, and our experiments next.
All of our experiments are conducted on a single NVIDIA A100-PCIE-40GB GPU.

\subsection{TempRALM Component Configuration}
% Already covered in Methodology, but still keeping this here for now
The TempRALM retriever uses the standard Atlas-large~\cite{izacard2022atlas} configuration, augmented with our temporal extensions.  In particular, it is based on a general-purpose dense retriever that uses a dual-encoder architecture, based on the Contriever \citep{contriever} and a sequence-to-sequence model built on the Fusion-in-Decoder architecture \citep{fusion_in_decoder} which uses T5-1.1 \citep{T5} with a language modeling adaptation. We initialize the retriever and generator with the same pre-training as Atlas \citep{izacard2022atlas}. 

 %For our experiments, we use Atlas-large which has 110 million parameters in the retriever, and 770 million in the LLM.

We chose parameters to configure TempRALM and Atlas-large based on experimenting with a range of values across their various hyper-parameters, including $top k$ documents to retrieve for every question, number of training steps, retriever and language model learning rates, and sampling temperatures.

\subsection{Document Index Generation}
We initially explored using Wikipedia \cite{wiki:} for use as a document index, as it is commonly used in knowledge intensive tasks.  However, we found that the structure of Wikipedia pages tends to make it harder to connect events to the dates on which they occurred due to a lot of additional text present on each page, resulting in poor model performance. In addition, once we crawled all Wikipedia pages and parsed the documents related to tennis grand slams, we found that the ATP/WTA Men's and Women's Singles Tournament Tennis data table \citep{atp:} actually contains was more detailed information for each event, such as who played which match, dates and locations of games.  
In fact, as tables are a common way to summarize details of events, we believe that exploring their use for training, fine tuning or prompting LLMs is an important direction to explore.  We take an initial step towards this goal in this paper.
 
We started out by filtering the ATP table by the following 4 grand slams: Australian Open, Roland Garros, Wimbledon and US Open. We then converted each row of the table into a passage of text described in natural language, covering tournament names, winner names, date of match and final scores.  We also extracted the timestamp from the row, in the format, $YYYY-MM-DD$  and attached it the passage as metadata. We span tournaments from 1978 to 2019, yielding to 50,000 documents. The process is illustrated in Figure \ref{Fig:CSV_to_passage_generation}. 

\begin{table*}[ht]
\centering

\begin{tabular}{@{}ccccc@{}}
\toprule
          & \multicolumn{2}{c}{TPQ-2019 Test Set Exact Match} & \multicolumn{2}{c}{TPQ-2020 Test Set Exact Match} \\ \cmidrule(l){2-3} \cmidrule(l){4-5} 
Train Set & Atlas              & TempRALM                    & Atlas                 & TempRALM                        \\ \midrule
32        & 64.84              & \textbf{71.72}              & 37.65                 & \textbf{55.93}                 \\
64        & 74.68              & \textbf{75.78}              & 40.93                 & \textbf{68.44}                 \\
128       & 76.88              & \textbf{77.80}               & 41.72                 & \textbf{72.65}                 \\ 
\bottomrule 
\hspace{2mm}
\end{tabular}
\caption{Comparing Temporal Proximity with Time based token matching: We evaluate the retrieval performance of two query sets - TPQ-2019 and TPQ-2020 - which contain identical questions about an event that occurred in 2019 but have timestamps from two consecutive years, to simulate the time-frame in which the queries are made. This table presents the performance comparison between the Atlas-large model and TempRALM. Both models were fine-tuned on a limited training sets comprising 32, 64, and 128 time-suffixed question-answer pairs. A robust model is expected to accurately address  queries based on their temporal closeness to passages relevant to the event, rather than relying solely on matching year tokens within the queries and passages. Our model demonstrates strong performance across both query sets.
}
\label{tab:Baseline_vs_Temporal_EM}
\end{table*}

%\subsubsection{Input Format}
\subsection{Training and Evaluation Data}

Since our focus is on temporally aware question answering, we use the following input data format for our model: (query, timestamp, answer). The time stamp captures the time at which a query is asked, and the answer corresponds to the answer which is valid at the time the query is posed.  

For instance, given the query "Who won the Australian Open Women's singles final?" posed on May 14, 2012, the correct response should indicate the player who held the title of Australian Open Women's singles champion as of May 14, 2012. It's crucial to note that the response to the same query changes as soon as a new Australian Open tournament completes.

\subsubsection{Few-Shot Training Set}

We create 3 non-overlapping few-shot training sets of size 32, 64 and 128 data points each. Each data point has the format (query, timestamp, answer). Each training set is constructed to provide a balanced distribution of query types, e.g., queries about winners, runners-up, finalists, and final match scores, spanning all four major tennis tournaments from 1978 to 2018. 

We chose the sizes of the few-shot learning training sets based on recommendations by the Atlas authors~\cite{izacard2022atlas}, who found that 64 training examples typically result in reasonable performance. We added 32 and 128-shot experiments in order to observe the impact of training set size in our specific scenario.

\subsubsection{Test sets}

We created two evaluation test sets, each of size 128, which use ground truth triplets using the same format and content as the training set, except focusing on tournament data for 2019, creating a clear separation between the training and test sets.  We refer to these test sets as Temporal Proximity Query sets (TPQs), namely TPQ-2019 and TPQ-2020. Both query sets contain identical questions pertaining to the matches that occurred in 2019 but have timestamps from two consecutive years, 2019 and 2020 respectively.  This enables us to simulate two temporal contexts. For instance, given the question 'Who won the US Open Women's singles final?', TPQ-2019 would time stamp the query as being posed in 2019, e.g., with 12-31-2019, whereas TPQ-2020 would  timestamp the query as being posed in 2020, e.g., 01-01-2020. 
The answer in both cases refers to the U.S. Open tournament which took place in September 2019. 

Answering the question correctly requires the model to go beyond textual pattern matching, and to have (or be able to simulate) temporal awareness that enables it to determine proximity between dates.
Figure \ref{Fig:Query_examples} shows how we constructed queries relevant to a particular passage. 

\section{Results}

\label{sec:results}

For our baseline experiment, we compare the performance of TempRALM to Atlas-large on  TPQ-2019 and TPQ-2020 test sets in 32, 64, and 128 few-shot training scenarios, using an exact match metric.  In addition, we evaluate the retrievers of both models using a recall@1 and recall@5 metrics, which compute the number of times the correct passage is in the top 1 or top 5 results returned by the retriever respectively. 
We discuss our results next.

% jj: removed the below because it may be repetitive; will revisit once the section is complete
% We performed an ablation study to assess the difference in the 2 retrieval strategies. We conduct 5 experiments for each Query set (TPQ1 and TPQ2) for both baseline and TempRALM method, in 32, 64, and 128 shot settings. We also explore the over-retrieval of passages for our temporal methodolody. To evaluate how well the retriever is performing, we calculate the recall@1, recall@5 and recall@20 metrics. We also note our observations, and limitations of our method.

\subsection{Comparison between TempRALM and Atlas-large}

We compare the performance of TempRALM and Atlas-large on both the TPQ-2019 and TPQ-2020 test sets, using an exact match metric (Table \ref{tab:Baseline_vs_Temporal_EM}). Each experiment was run 5 times and the results in the table are the average of the 5 runs. 
We observe that when the timestamp year of the query matches that of the event (and thus the date referenced in the corresponding text passage), the unmodified Atlas model performs comparably to TempRALM.  In these scenarios, the semantic score captures the matching date pattern between the query and the passage.
%We observe that TempRALM is equivalent to baseline performance TPQ-2019, where questions are asked in the year 2019. However, it significantly improved upon performance in TPQ-2020 query set, where questions are asked in the year 2020. 
%We observe that the baseline retrieval performs well when the year in the query matches the year in the passages. It scores the passages a higher semantic score (due to the token matching). 
However, when the year in the timestamp of the query does not match that in the text passage (TPQ-2020 experiments), TempRALM outperforms unmodified Atlas by 49\% in the 32 few-shot learning experiment, 67\% in the 64 few-shot learning experiment, and 74\% in the 128 few-shot learning experiment.  This illustrates that our temporal augmentation plays a key role in the overall answer generations since the the year in the query and passages no longer match, and the time proximity between different time stamps is not captured by the similarity score, but only by our temporal extensions. 

Our findings also indicate that as the number of examples in the few shot training increase, our temporal augmentation becomes increasingly more impactful.

Similarly to the findings in \citep{izacard2022atlas}, we find that at few-shot training with 64 examples, model performance starts to be reasonable, and improves slowly for 128 training examples. We show the Exact match performance in Table \ref{tab:Baseline_vs_Temporal_EM}

% It fetches passages relevant to the tournament, but fetches older passages, and gets confused by appending timestamp.
%TempRALM had a 74\% improvement over baseline, suggesting that it can be implemented in retrievers to fetch relevant content when there are multiple versions of relevant content for different points of time present together in the index. 

%\vspace{-5mm}

%\subsection{Comprehensive coverage with over retrieval}
%We over retrieve passages surpassing the $top k$ threshold to provide the retriever with a broader coverage and thereby reinforce its capacity to fetch passages that are temporally closer to the query. To illustrate, if the $top k$ is set at 20, we conducted experiments with varying levels of over-retrieval: none, +40, +80, +100, and +120. Our findings indicate that increased over-retrieval yields superior results. However, due to constraints in computational resources, we limit our over-retrieval to +100. We show the impact of over-retrieval in 32-shot setting in Figure \ref{fig:Over-retrieval}

\subsection{Retriever performance analysis}

We observed instances where TempRALM successfully retrieves the gold passage (which contains the correct answer), but the answer generated by the LLM is wrong, indicating that the problem in those examples lies with the LLM. 
%For instance, in a specific case, the retriever correctly identified and ranked the relevant passage as highest for the question, which stated that Roger Federer was the runner-up in the Wimbledon Men's Singles final. However, the generated response provided 'Andy Murray' instead. 

% example - 
% query: Who was runner-up in the Wimbledon men's Singles final? timestamp: 2020-01-01, answers: Roger Federer, generation: Andy Murray, passages: [ 'Novak Djokovic .. defeated Roger Federer .. at the Wimbledon Men's .. on 2019-07-01, in the final match ..] 

To quantify this behavior, we calculated the number of times the retriever ranked the gold passage as the top retrieved document (recall@1) and also the number of times the retriever ranked the gold passage in the top 5 retrieved passages (recall@5) for both TempRALM and Atlas-large. Our results, shown in Table \ref{tab:recall_metrics}, illustrate that temporal retrieval improves both Recall@1 and Recall@5, and this improvement is the most pronounced in the Recall@1 for the TPQ-2020 experiments, where TempRALM improves over Atlas-large by a factor of 165\%.  These experiments illustrate both the inability of Atlas-large's retriever to handle temporal data, and TempRALM's impressive ability to compensate for that.

% \begin{table}[ht]
% \centering
% \begin{tabular}{@{}lllll@{}}
% \cmidrule(r){1-4}
% Model    & Test Set & R@1 & R@5 &  \\ \cmidrule(r){1-4}
% Atlas    & 2019     & 69  & 80  &  \\
% Atlas    & 2020     & 31  & 80  &  \\ \cmidrule(r){1-4}
% TempRALM & 2019     & 80  & 96  &  \\
% TempRALM & 2020     & 82  & 96  &  \\ \cmidrule(r){1-4}
% \end{tabular}
% \caption{}
% \label{tab:recall}
% \end{table}
\begin{table}[h]
\centering
\begin{tabular}{@{}l|ll|ll@{}}
\toprule
         & \multicolumn{2}{l|}{TPQ-2019} & \multicolumn{2}{l}{TPQ-2020} \\ \midrule
Model    & Recall@1         & Recall@5         & Recall@1         & Recall@5        \\ \midrule
Atlas-Large    & 0.54          & 0.63          & 0.24          & 0.63         \\
TempRALM       & 0.63          & 0.75          & 0.64          & 0.75         \\ \bottomrule
\end{tabular}
\vspace{1em}
\caption{Retriever Recall Metrics: We calculate the recall in 64-shot training example setting. We choose an experiment closest to the average exact match of all our 64-shot Atlas and TempRALM experiments.}
\label{tab:recall_metrics}
\end{table}
% \vspace{-5mm}

\subsection{Model Size Performance: Large vs Base Model}
\label{atlas-large-comparison}
We evaluated the performance on the Atlas large model that has (reader 770M / retriever 110 M) Parameters. We also compared performance on the Base ( Reader 220M / retriever 110M) model, and show that the temporal retrieval works better across different model sizes. In table \ref{tab:base_vs_large}, we compare the Exact Match performance of a 64-shot experiment setting experiment for Atlas and TempRALM. 

\begin{table}[h]
\centering
\begin{tabular}{@{}ccccc@{}}
\cmidrule(r){1-3}
               & \multicolumn{2}{c}{Test Set Exact Match} &  &  \\ \cmidrule(lr){2-3}
Model Size     & TPQ-2019            & TPQ-2020           &  &  \\ \cmidrule(r){1-3}
Atlas Base     & 42.97               & 28.13              &  &  \\
TempRALM Base  & 56.25               & 35.94              &  &  \\ \cmidrule(r){1-3} 
Atlas Large    & 74.68               & 40.93              &  &  \\
TempRALM Large & 75.78               & 68.44              &  &  \\ \cmidrule(r){1-3}
\end{tabular}
\caption{Model Size Performance: We compare the Exact Match in 64-shot training example setting between Atlas and TempRALM across base and large models.}
\label{tab:base_vs_large}
\end{table}

%\subsection{Limitations of our method}

%\begin{itemize}
%    \item We've  observed situations where, in cases of passages sharing exact dates, the semantic score becomes the decisive factor in determining the final score. This occurs because the temporal score remains identical for both passages. This could have implications for architectures that implement periodic or batch ingestion of passages, potentially assigning the same date to all documents within a given batch.

%\end{itemize}

\section{Conclusions and Future Work}

\label{sec:conclusions}
 
In this study, we introduced and evaluated TempRALM, a Retriever Augmented Language Model (RALM) augmented with temporal awareness. Unlike conventional RALM approaches that rely solely on semantic similarity, TempRALM considers both semantic and temporal relevance when selecting documents to pass to its Large Language Model (LLM) in response to a given query. Our results indicate an improvement in performance of up to 74\% compared to the Atlas-large model, even when multiple versions of documents (from different time points) are present in the document index. Notably, we achieve this without the need for model pre-training, replacing the document index with an updated index, or adding any of other computationally intensive elements. We plan to explore a number of avenues for building on the work presented in this paper, such as implementing and evaluating different learning strategies for the parameters of our temporal relevance function, and exploring the interplay between the retriever and LLM. Furthermore, we plan to explore the use of our temporal retrieval approach in other tasks such as fact checking, recommender systems, and retrieval augmented dialog agents. 
%This technique can easily be extended to other dimensions such as geographical locations and cross-modal retrieval.

%\section{References}

% \bibliographystyle{ACM-Reference-Format}
\bibliography{anthology}

\bibliographystyle{ACM-Reference-Format}

%\appendix


\end{document}